\documentclass[letterpaper, 10 pt, conference]{IEEEtran}  

\IEEEoverridecommandlockouts 

\usepackage{epsfig} 
\usepackage{amsmath} 
\usepackage{epstopdf}
\usepackage{amsmath} 
\usepackage{amssymb}  
\usepackage{epstopdf}
\usepackage{algorithm}
\usepackage{algpseudocode}
\usepackage{color}
\usepackage{xcolor}

\usepackage{url}
 \usepackage{subfigure}

\title{\LARGE \bf
A Tube-based Robust MPC for a Fixed-wing UAV: an Application for Precision Farming
}

\author{M. Mammarella$^{1}$, E. Capello$^{2}$ 
\thanks{$^{1}$M. Mammarella is with the Department of Mechanical and Aerospace Engineering, Politecnico di Torino, Corso Duca degli Abruzzi, 24, 10129 Torino, Italy, {\tt\small martina.mammarella@polito.it
}}
\thanks{$^{2}$E. Capello is with the Department of Mechanical and Aerospace Engineering, Politecnico di Torino and with the CNR-IEIIT, Politecnico di Torino, Corso Duca degli Abruzzi 24, 10129 Torino, Italy, {\tt\small elisa.capello@polito.it}}%
}

\begin{document}

\maketitle
\thispagestyle{empty}
\pagestyle{empty}

\begin{abstract}
The techniques of precision agriculture include the possibility to execute crop monitoring tasks through the application of Unmanned Aerial Vehicles (UAVs). These platforms are flexible, easy to use and low-cost, and they are the best candidates for improving the farm efficiency and productivity. In this research, a guidance algorithm and a robust control system are combined to guarantee the robustness of the system to additive noise (i.e. wind disturbance) and uncertainties (i.e. model parameter variations). A small fixed-wing UAV with an autonomy of about $1$ hour is proposed as case study, to reduce the cost of monitoring and increasing the stability performance of the system. A waypoint-grid on a paddy field is verified by hardware-in-the loop tests. The control scheme provides good results with a low computational effort, guaranteeing the repeatability of the monitoring and reduction of the costs.

\end{abstract}

\section{INTRODUCTION}
Autonomous unmanned aerial vehicles (UAVs) have
been increasingly used by civilian applications and, more recently, the most demanding applications are related to the precision agriculture sector, due to the great benefits given by the use of these platforms and the tools they incorporate, what provides the farmer with useful information. As clearly explained in \cite{pederi}, the key problem in precision farming and, in particular, in crop protection methods is the lack of efficiency and flexibility. Current methods are based on land machines, that are slow and expensive, compared to UAVs. Moreover, UAV technologies can provide real-time data to farmers, including crop monitoring.

The main objective of this paper is to propose a low-cost and high-throughput method for a crop monitoring system which uses a fixed-wing UAV as an operating platform. The key feature of the proposed approach is the design of guidance and control algorithms able to perform the desired mapping, guaranteeing robustness of the system to uncertainties and disturbances, reducing the flight time and optimizing the path, for crop monitoring. Moreover, the use of a fixed-wing UAV
is justified to cover a huge terrain extension in a single flight, gathering the information previously said in a much more efficient way. These information can be post-processed or processed in real time to obtain an operation map. Thus, the UAV can incorporate in its system the designed map and optimally distribute, through “capsule dropping” techniques, seeds, fertilizer and so on.

Additionally, the guidance segment assumes a relevant role for the accomplishment of the UAV mission, providing a feasible trajectory, combined with the control system. In the last couple of years, different studies on application of UAVs can be found in literature. In \cite{ni}, a crop-growth monitoring system is proposed, in which a multi-rotor UAV is used as platform. The focus of \cite{ni} is on the sensor resolution and the results are proposed in terms of evaluation of Normalized Difference Vegetation Index (NDVI), but not in terms of UAV performance and capability to perform the desired mission. Moreover, as in \cite{ghazal}, a small amount of field is monitored, increasing the mission/mapping cost.
Since, as highlighted in \cite{voug}, in the context of precision farming, an accurate position control and a field coverage by waypoints is required, to minimize the maximum deviation between the UAV path and the desired one, \cite{voug} a nonlinear Model Predictive Control (MPC) is proposed as algorithm for the path tracking but the real application and feasibility of the algorithm is not taken into account.

Our idea is to design a software of an autopilot, which can be used for different tasks, in particular aerial mapping and evaluation of the vegetation index of a specified area. The proposed software is able not only to maximize the performance of the on-board sensors, but also to improve the stability of the platform. These results are obtained thanks to the combination of a path planner and a controller, able to guide the UAV in flight with no human assistance. The path-following control of the UAV can be separated into different layers: (i) inner loop for pitch and roll attitude, and airspeed control, (ii) outer loop on heading, altitude for the waypoints tracking, and (iii) waypoint navigation. The proposed software is verified by simulations and Hardware-In-the-Loop (HIL) tests for aerial mapping of a paddy field, with a multispectral sensor.

Moreover, the second objective is to validate the effectiveness of the combination of the guidance algorithm with a robust MPC for fixed-wing UAVs, focusing on the real-time feasibility of the proposed strategy, including both atmospheric disturbances (i.e. additive noise), model uncertainties (variations on speed $V$ and mass $m$) and platform inaccuracies (variations on the moments of inertia $J$). A Tube-based Robust MPC (TRMPC) \cite{cannon,raw} is proposed. This novel approach focuses on two main goals: (i) to provide robustness to disturbances and (ii) to maintain the computational efficiency of classical MPC strategies. The robustness of the controller can guarantee the repeatability of the path and the reduction of the cost. Moreover, the same controller can be used for different UAVs (fleet of UAVs), with similar but not identical characteristics, that can cover a bigger area with reduction of the cost and with different payload (i.e. sensors). 

Finally, the third objective of this paper is to evaluate the effectiveness of the proposed algorithms via HIL tests, which deal with reproducing the environment where the embedded system will run. This is
usually one of the last steps in the testing procedure, 
before integration and system tests. Furthermore,
rigorous experiments with HIL simulator reduce the risk of
damaging the equipment during further flight tests \cite{barros}.
With this motivation, we
performed HIL simulations to validate the guidance and
control algorithms, for a snake-based grid on a paddy field.

The novelty of the proposed approach is the combination of the proposed guidance and control algorithms, which can reduce the time to flight and optimize the monitoring of the selected area, guaranteeing robustness to disturbances and uncertainties. 
The advantages of the proposed approach are: (i) the ability to monitor with UAVs huge extension fields (1-10 ha paddy fields), larger than the ones tractors can cover, (ii) the increase of yield (about 20-30\%) applying precision farming methodology, and (iii), focusing on a paddy field monitoring, the capability of the proposed software segment to reduce the amount of fertilizers, the rice disease and waste of resources.

The paper is organized as follows. The precision farming scenario is described in Section \ref{scenario}. In Section \ref{model} the aircraft model and the on-board sensor for crop monitoring are described. In the same Section, the autopilot and HIL board are presented. In Section \ref{tube} the TRMPC  theory is described. Simulations and HIL tests are analyzed in Section \ref{results}. Finally, conclusions are drawn in Section \ref{conc}.

\section{Precision Farming and Agriculture Scenario}
\label{scenario}
As defined in \cite{pederi}, precision agriculture is the understanding of the complex interactions between crop growth and decision-making. With the help of precision agriculture it is possible to identify problematic areas and apply chemicals only to those areas. This will allow for significant savings.
For this reason, UAVs have several applications and purposes in precision agriculture. They can perform such tasks:
\begin{itemize}
\item NDVI Monitoring,
\item Plants Pathology Monitoring,
\item Crop Water Stress Index (CWSI) Monitoring,
\item Spraying of Liquid Fertilizers, Pesticides and Spraying,
of Entomological Material (Trichogramma), and
\item Aerial Mapping.
\end{itemize} 

\begin{figure}[thpb]
\centering
\includegraphics[width=7cm]{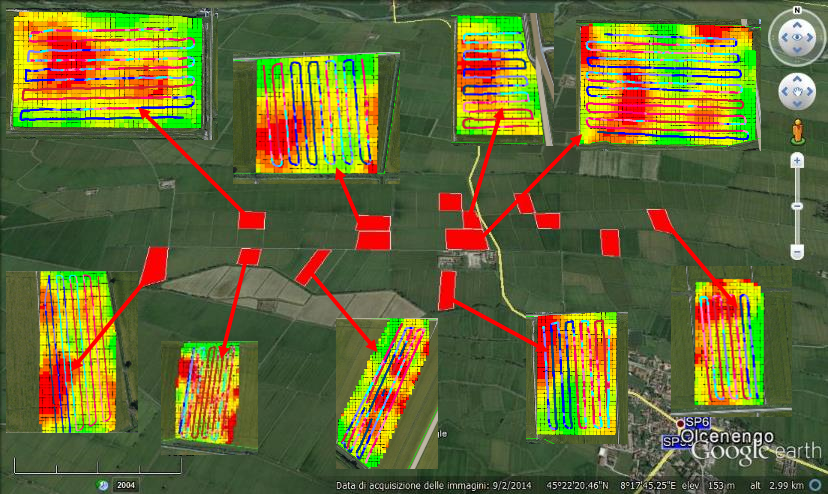}
\caption{Example of paddy field and snake grid \cite{smile}.}
\label{fig:scenario}
\end{figure}

This paper focuses in two tasks: (i) the guidance and control algorithms are able to perform an efficient and optimzed path to perform aerial mapping on the selected area, and (ii) a paddy field is monitored (with a fixed-wing UAV), in which a multispectral camera is installed on board, providing information about NDVI.
Precision agriculture allows farmers to know vegetation index in their crops, like the hydric stress level and the NDVI. The combination of aerial photographs of terrain with multispectral and IR cameras on the UAV are used to get the NDVI. This index is used for the evaluation of an intensity radiation of some bands on the electromagnetic spectrum emitted or reflected by radiation. Through this evaluation, farmers are able to know the evolution, quantity and quality of production in their crops.
An example of the analyzed area is in Figure \ref{fig:scenario}, in which different sub-areas and a snake-based grid are defined. 

\section{Mathematical Model and Hardware configuration}
\label{model}
In this section, the proposed fixed-wing UAV and the on-board sensor for paddy field monitoring are presented. Moreover, the autopilot and the HIL board are described. 
Finally, a brief description of the guidance algorithm is provided, since the novelty of the proposed approach is based on the control algorithm (deeply presented in Section \ref{tube}).

\subsection{Aircraft Model}
\begin{figure}[thpb]
\centering
\includegraphics[width=5cm]{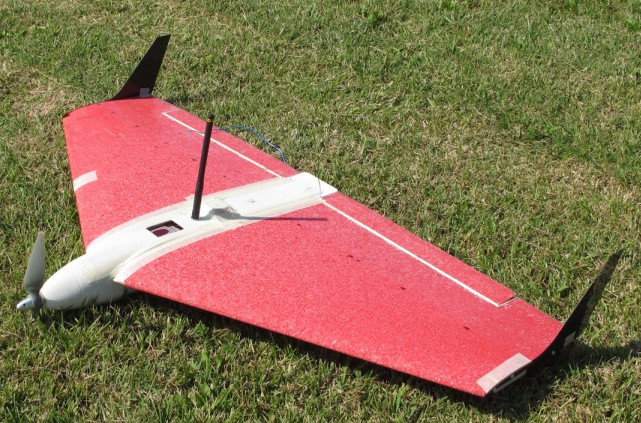}
\caption{The MH850 mini-UAV.}
\label{MH850}
\end{figure}

The aircraft considered for the controller implementation is the MH850 mini-UAV (\cite{cap2012}, \cite{cap13}). The MH850 has a tailless configuration, electric propulsion, and tractor propeller (see Figure \ref{MH850}). The wingspan is 85 cm, the approximate mass 1 kg, it is able to fly for about 45 minutes at a cruise speed of 13.5 m/s. Aircraft control is achieved with trailing edge elevon (symmetric deflection for elevator $\delta_e$ and antisymmetric for aileron $\delta_a$). 
A database including all the aerodynamic derivatives is employed to design the linear and nonlinear aircraft models \cite{cap2008,cap2012}.

Reference flight conditions for the model are speed $V_0=13.5$ m/s, altitude $h_0=100$ m, angle of attack $\alpha_0=5.18$ deg and $\theta_0=5.18$ deg. The equations of motion linearization procedure results in the decoupling of the longitudinal and lateral-directional planes. Each of them is modeled with standard continuous time-invariant state space representation

\begin{equation}
\begin{aligned}
&\dot{x}(t)=Ax(t)+Bu(t),\\
&y(t)=Cx(t),
\label{state_space_c}
\end{aligned}
\end{equation}
where $x(t)$ is the state vector, $u(t)$ the control signal, $y(t)$ the controlled output, $A$ the state matrix, $B$ the input matrix and $C$ the output matrix. Matrices $A$, $B$ and $C$ are built according to \cite{stevens}, the aerodynamic derivatives in the matrices are obtained by a validated software based on the extended lifting-line theory.

The longitudinal state variables are the airspeed along the X axis $u$, the angle of attack $\alpha$, the pitch angle $\theta$ and the pitch rate $q$. The control inputs are the throttle $\Delta T$ and the elevator deflection $\delta_e$.
The lateral-directional state variables are the lateral airspeed $v$, the roll rate $p$, the yaw rate $r$ and the roll angle $\phi$ and the yaw angle $\psi$. The only input is the aileron deflection $\delta_a$. 
The controller parameters, tuned for the decoupled linear model, are validated considering a complete nonlinear model obtained from the aircraft equations of motion as defined in \cite{etk}. These are a set of $12$ equations describing the forces, moments, angles and angular speeds which characterize the flight condition of the aircraft. Trim conditions coincide with the equilibrium values used for the linear model, previously described.

\subsection{On-board Sensor}
The active sensor considered in this work is the OptRx$\textsuperscript\textregistered$ crop sensor from AG Leader (www.agleader.com). This sensor provides immediately the VI required. A data logger elaborates the data acquired by the OptRx$\textsuperscript\textregistered$ crop sensor and associates these readings to the onboard RTK based GPS  to create
georeferenced maps. 

\subsection{Autopilot and Hardware in the Loop}
\begin{figure}[h!]
\centering
\subfigure[The autopilot board]{\includegraphics[width=0.49 \columnwidth]{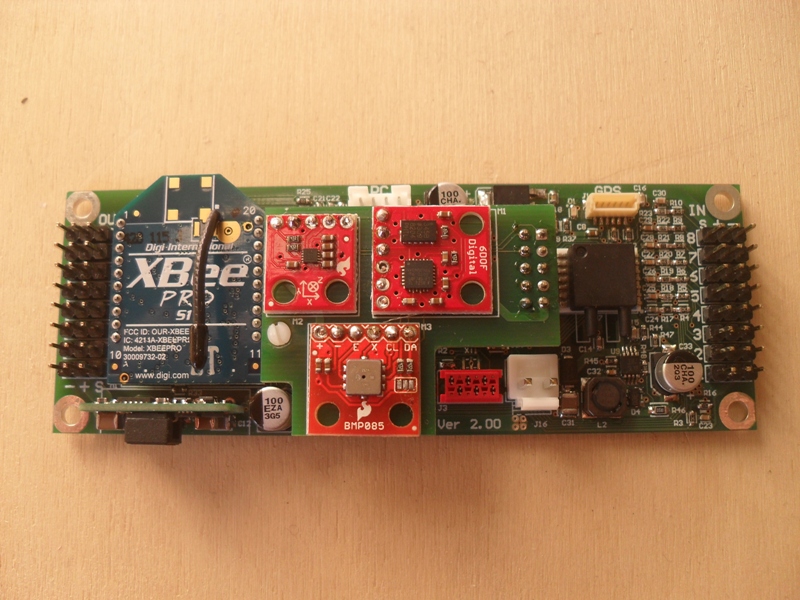}%
\label{autopilot}}
\hfil
\subfigure[Cable connection of XMOS board]{\includegraphics[width=0.49 \columnwidth]{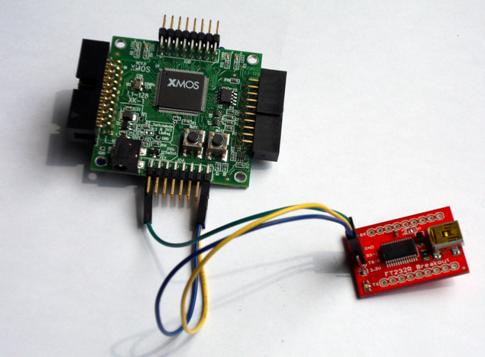}%
\label{XMOS}}
\caption{Hardware configuration considered.}
\label{HW}
\end{figure}

A custom-made autopilot is installed on-board and was designed and produced in the Department of Mechanical and Aerospace Engineering of Politecnico di Torino \cite{cap2012} (see Figure \ref{autopilot}). Main characteristics comprehend an open architecture and the possibility to be reprogrammed in flight and real time telemetry. Sensors include GPS, barometric sensor, differential pressure sensor and three-axis gyros and accelerometers. The CPU is the ATXMEGA256A3U-3U model with 256Kb flash memory and 16Kb of RAM. A Radiomodem Xbee Pro S1 is used for the communication link between the Ground Control Station (GCS) and the autopilot.


To validate the controller and the guidance algorithm with
HIL simulations, a commercial board (the XMOS XK-1A
board) with USB connection has been selected because of its
characteristics and potentialities (i.e. flash memory of 128 Kb
and a CPU clock of 20 MHz) similar to our microcontroller.
This board must be physically connected to the laptop to
effectively be part of the loop.
The XMOS XK-1A is a low-cost development board
produced by XMOS Ltd (www.xmos.com), and it is
characterized by the multi-core multi-thread processor
XS1-L1 which is able to perform several real-time tasks. Its
parallel computing ability is essential for unmanned
applications where high level tasks (for instance the control
logic) have to be combined with low level assignments (such as
I/O) \cite{martins}. 
One of the main advantages in using XMOS technology is
the facility in programming the board. The language for the
XMOS board is called XC, which can be compared with C language, and it shows some additional commands for the
management of the ports and the pins.
Focusing on the HIL simulation, the inner and outer loop
controllers are partially converted in XC language and
implemented in the XMOS platform. To accomplish this connection, a breakout board for USB to serial conversion is placed between the XMOS board
and the laptop. A detail of the HIL cables connection is
represented in Figure \ref{XMOS}.

\subsection{Guidance Algorithm}
The guidance algorithm here proposed is deeply described in \cite{cap2014}, in which some simplifying hypotheses, according to the flash memory limitation of the autopilot microcontroller, are taken into account. A given set of
waypoints is considered, with assigned North, East and
altitude coordinates. This set of waypoints includes the
starting point, that is the point where the UAV finishes the
climb and the autonomous flight starts. The starting point and
all the waypoints are supposed to be at the same altitude; thus,
a 2D path is considered. A trajectory smoother, that makes cinematically feasible the assigned trajectory in terms of speed and turn rate
constraints, is implemented.
For the evaluation of the performance of the guidance algorithm for aerial mapping, the Cross-Track Error (CTE) $\epsilon_r$ is calculated, to monitor the UAV position
with respect to the reference path. We consider the UAV
real position $P_{UAV}$ in terms of UAV East and North
coordinates, respectively, $E_{UAV}$ and $N_{UAV}$ and the
segment connecting two waypoints in terms of previous waypoint
WP$_n (E_n, N_n)$ and next waypoint WP$_{n+1}(E_{n+1}, N_{n+1})$.
The cross-track error is then calculated as
\begin{equation}
\label{eq:cte}
\epsilon_r = \frac{\vert E_{UAV}-mN_{UAV}-(E_n-mN_n)}{\sqrt[]{m^2+1}} ,
\end{equation}
with $m=\frac{E_{n+1}-E_n}{N_{n+1}-N_n}$.

\section{Tube-based Model Predictive Control}
\label{tube}

Let consider the discrete-time state space formulation of \eqref{state_space_c}

\begin{equation}
\text{x}_{k+1}=A_d\text{x}_k+B_d\text{u}_k+\text{w}_k,\\
\label{state_space}
\end{equation}
where $\text{x}_k$ is the state vector, $\text{u}_k$ the control signal, $\text{w}_k$ the unknown bounded uncertainty, $A_d$ the discrete state matrix and $B_d$ the discrete input matrix. The system is required to satisfy the state and input hard constraints 

\begin{equation}
\text{x}_k\in\mathbb{X}, \,\,\,\, \text{u}_k\in\mathbb{U},\\
\label{eq:set}
\end{equation}
where $\mathbb{X}\subset\mathbb{R}^n$ and $\mathbb{U}\subset\mathbb{R}^m$ are compact and convex polytope containing the origin. Moreover, the noise $\text{w}_k$ is a realization of a stochastic process, each one an independent and identically distributed (i.i.d.) zero-mean random variable, with a convex and bounded support $\mathbb{W}\subset\mathbb{R}^n$. 

The TRMPC approach is based on the concept of \textit{tube} of state trajectories, each one representing an admissible disturbance sequence $\textbf{w}$ over the observed time-window. The center of this tube is represented by the nominal undisturbed dynamics

\begin{equation}
\text{z}_{k+1}=A_d\text{z}_k+B_d\text{v}_k,\\
\label{und_state_space}
\end{equation}
where $\text{z}_k$ and $\text{v}_k$ represent the nominal state and input respectively. 

The TRMPC allows to steer the uncertain trajectories to the nominal one, controlling  the "center" of the tube via a classical MPC approach. In order to ensure the robustness of the algorithm, the constraint set imposed on the nominal system are tightened in compliance with the guidelines provided in \cite{raw}, as a function of the minimal robust positive invariant set for $\text{x}_{i+1|k}=A_d\text{x}_{i|k}+\text{w}_{i|k},\,\,\text{w}\in \mathbb{W}$.

Moreover, to stabilize the system with respect to parametric uncertainty $\text{q}$, ascribable for example to discrepancies between the mathematical model and the actual dynamics, neglected nonlinearities and manufacturing process, a Linear Matrix Inequality (LMI) approach applied to the definition of the Schur stability of the closed-loop system has been considered, obtaining the feedback gain matrix $K$ that robustly stabilizes the system
\begin{equation}
\text{x}_{i+1|k}=(A_d+B_dK)\text{x}_{i|k}+B_d\text{v}_{i|k}+\text{w}_{i|k}.
\label{closed_loop}
\end{equation}

Then, following the typical approach adopted for classic MPC, the finite horizon optimal quadratic cost can be defined for the nominal system, instead of the uncertain system, as

\begin{equation}
J_N(\text{z}_k,\textbf{v}_k)=\sum_{i=0}^{N-1}(\text{z}_{i|k}^TQ\text{z}_{i|k}+\text{v}_{i|k}^TR\text{v}_{i|k})+\text{z}_{N|k}^TP\text{z}_{N|k},
\label{nom_cost}
\end{equation}
where $\textbf{v}_k$ represents the control sequence over a $N$-step prediction horizon. $Q\in\mathbb{R}^{n\times n}$, $Q\succ0$, and $R\in\mathbb{R}^{m\times m}$, $R\succ0$ are the state and control weight matrices whereas $P\in\mathbb{R}^{n\times n}$ is the terminal one, solution of the discrete Algebraic Riccati equation. Hence the nominal finite horizon optimal control problem is defined as

\begin{subequations}
\begin{equation}
\begin{aligned}
& \underset{\textbf{v}_k}{\text{min}}
& & J_{N}(\text{z}_k,\textbf{v}_k) 
\end{aligned}
\end{equation}
\begin{equation}
\begin{aligned}
& \text{s.t.}
& & \text{z}_{i+1|k} = {A}_d\text{z}_{i|k}+{B}_d\text{v}_{i|k},\quad \text{z}_{0|k}=\text{x}_{k},\\
&
& & \text{z}_{i|k}\in \mathbb{Z}, \quad i\in [1,N-1],\\
& 
& & \text{v}_{i|k}\in \mathbb{V}, \quad i\in [0,N-1],\\
& 
& & \text{z}_{N|k}\in \mathbb{Z}_{f}.
\end{aligned}
\end{equation}
\label{eq:mpc_nom}
\end{subequations}
The first control action $\text{v}_{0|k}^*$ of the optimal sequence $\textbf{v}_k^*$, solution of \eqref{eq:mpc_nom}, represents the optimal control applied to the nominal system. The correspondent control on the uncertain system is obtained applying a duel-mode prediction scheme 

\begin{equation}
\text{u}_k=\text{v}_{0|k}^*+K(\text{x}_k-\text{z}_k).
\label{u_k}
\end{equation}
The final TRMPC algorithm can be summarized as shown in Algorithm \ref{Online_alg}.

\begin{algorithm}
\caption{TRMPC Algorithm}
\label{Online_alg}
\begin{algorithmic}[1]
\Procedure{ }{}
\State \textit{Offline}: Evaluate the feedback gain matrix $K$ and the nominal constraint sets $\mathbb{Z}$ and $\mathbb{V}$. Set N.
\State \textit{Online}: At current time $k$, evaluate $x_{i=0|k}=x_{k}$. 
\For{$i=0:N-1$ }
\State Set $\text{z}_{i=0|k}=\text{z}_{0|k}=\text{x}_{k}$
\State Solve \eqref{eq:mpc_nom}
\EndFor
\State Get $\textbf{v}_{0}^*$ and extract the first control action $v_{0}^*$.
\State Evaluate $u_{k}$ according to \eqref{u_k}, then evaluate $x_{k+1}$ applying $u_{k}$ on \eqref{state_space}.
\EndProcedure
\end{algorithmic}
\end{algorithm}

\section{Simulation and HIL Results}
\label{results}
\begin{figure}[h!]
\centering
\includegraphics[width=1\columnwidth]{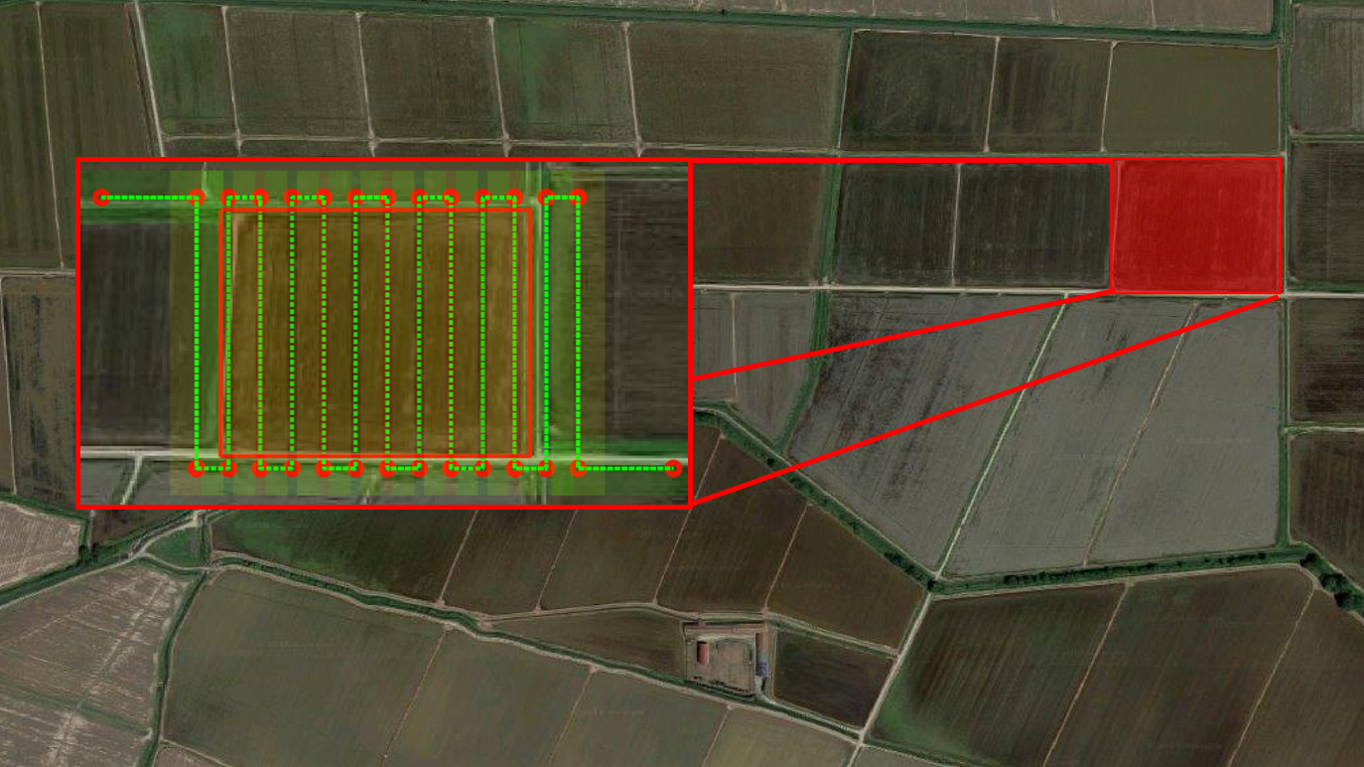}
\caption{Monitored portion of field at Olcenengo, Italy, overlapped with a snake-path.}
\label{fig:f101}
\end{figure}

The case-study chosen to validate the guidance and control strategy proposed in this work is a paddy field at Olcenengo, Vercelli, Piedmont, Italy ($45^{\circ}22'22.2''\text{N}, 8^{\circ}17'34.3''\text{E}$). The UAV flight mission is represented by a snake-path identified through a series of waypoints over a $200$X$150$ m rectangular-shape area. The Ground Sampling Distance (GSD) requirement is a function of the flight altitude, as described in \cite{mascarello2017feasibility}, and defines the grid width. In particular, in compliance with the recommendations of the multispectral camera manufacturer, for an altitude of $100$ m, the grid size has been set to $20$ m and is defined including a $10\%$ of both overlap and sidelap requirements, identified in Figures \ref{fig:f101} by light-yellow/green strips. Moreover, it is possible to notice how the coverage area includes also an additional $10$ m band for the flight path in order to allow the stabilization of the UAV for a straight flight after each turn. 

Both Software-In-the-Loop (SIL) and HIL simulations have been performed to validate the guidance and control strategy, considering the following initial conditions: (i) altitude $h_0=100$ m; (ii) airspeed $V_0=13.5$ m/s; (iii) angle of attack $\alpha_0=5.18$ deg; (iv) ramp angle $\gamma_0 = 0$ deg. The simulations and MPC parameters adopted in all the simulations are reported in Table \ref{t:mpc_param}. The heading angle is controlled by a Proportional Integrative Derivative (PID) control system. The results presented in the following have been obtained exploiting a simulator developed with MATLAB/Simulink 2016b.


\begin{table}[!h]
\renewcommand{\arraystretch}{1.3}
\caption{Simulations and TRMPC parameters adopted for both SIL and HIL.}
\label{t:mpc_param}
\centering
\begin{tabular}{c c}
\hline \hline
Parameter & Value\\
\hline
System sample time [s] & $0.01$\\
Board/PID sample time [s] & $0.05$\\
TRMPC sample time [s] & $0.1$\\
Prediction horizon [-] & $15$\\
$diag(Q_{long})$ & $[10^6,4\times 10^1,4\times 10^1,4\times 10^1,10^5]$\\
$diag(R_{long})$ & $[4\times 10^2, 3\times 10^{-6}]$\\
$diag(Q_{lat})$ & $[10^1,10^1,10^1,10^4]$\\
$R_{lat}$ & $10^6$\\
\hline \hline
\end{tabular}
\end{table}

For what concerns the disturbances introduced in the model, a $10\%$ variation over the airspeed, mass and inertia has been considered whereas the additive noise represents the presence of a persistent wind gust, modeled as a random samples with uniform distribution and a maximum intensity of 0.01 m/s in both longitudinal and lateral directions. Moreover, all the other state variables are supposed to be affected with a smaller impact, obtaining two different disturbance sets, $\mathbb{W}_{long}=[w_u,w_{\alpha},w_{\theta},w_{q},w_{h}]=[10^{-2},10^{-6},10^{-6},10^{-6},10^{-3}]$ for the longitudinal dynamics, and for the latero-directional $\mathbb{W}_{lat-dir}=[w_v,w_{p},w_{r},w_{\phi}]=[10^{-2},10^{-6},10^{-6},10^{-6}]$. 

States and controls constraints have been imposed, among which the more stringent ones are related to the command saturation of the elevator and aileron surfaces, limited to $\pm25$ deg. 

\begin{figure}[h!]
\centering
\includegraphics[width=0.85\columnwidth]{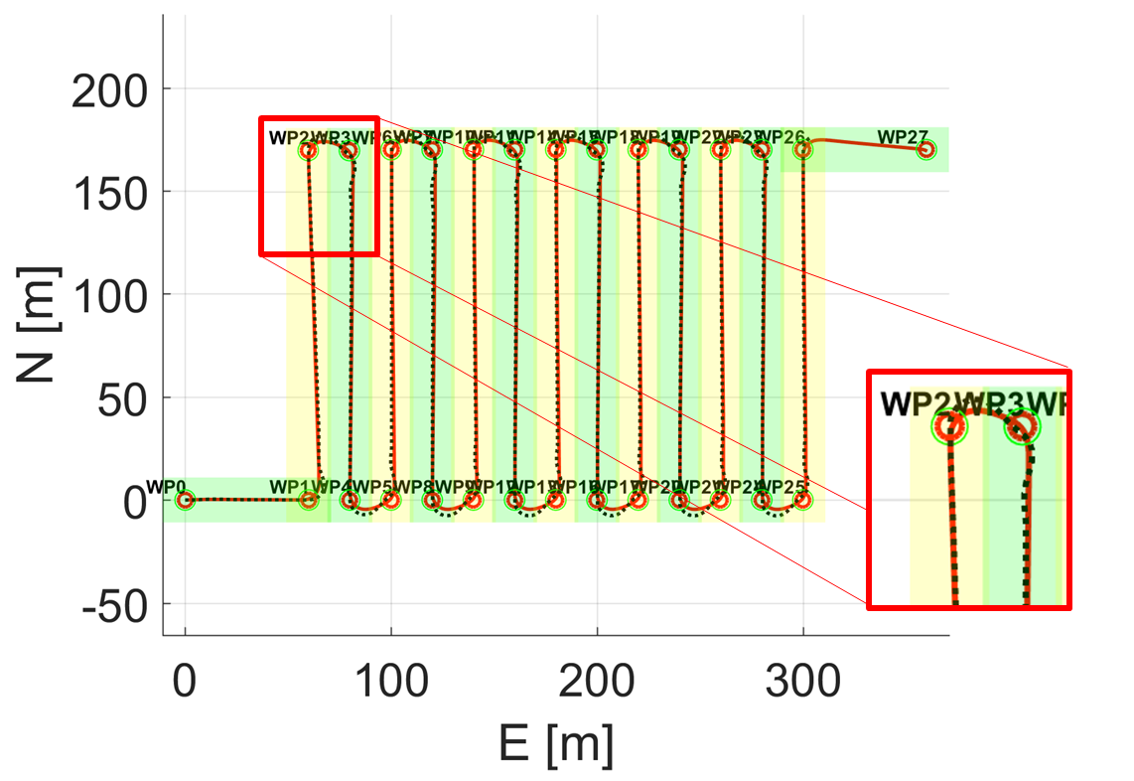}
\caption{Comparison among SIL (red line) and HIL (dotted black line) trajectories.}
\label{fig:f102}
\end{figure}

Figure \ref{fig:f102} depicts the trajectories obtained with SIL and HIL simulations. Marginal discrepancies can be observed among the two trajectories, mainly close to the turning point before the straight lines. This is due to the delay, inherent to the XMOS board.


\begin{figure}[h!]
\centering
\includegraphics[width=1\columnwidth]{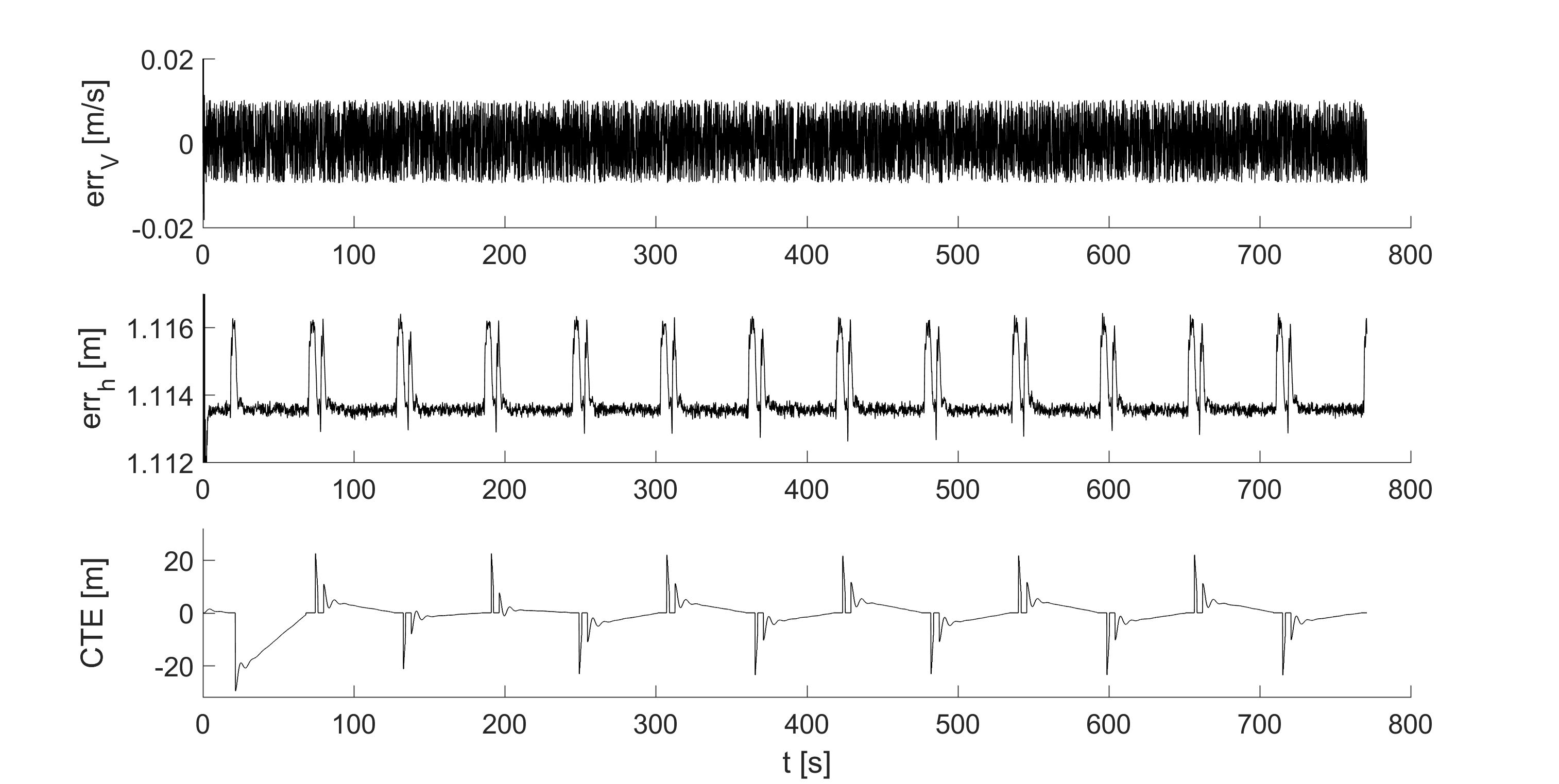}
\caption{Deviations between actual airspeed and quote with respect to the references, and cross-track error for the UAV position.}
\label{fig:f104}
\end{figure}

Moreover, in Figure \ref{fig:f104}, the deviations of airspeed and altitude, $err_V$ and $err_h$ respectively, with respect to references, i.e. $V_{ref}=V_0$ and $h_{ref}=h_0$, as well as the CTE error (Eq.~\ref{eq:cte}) are represented. We can observe that the airspeed is more noisy with respect to the altitude, due to the external disturbances.

\begin{figure}[h!]
\centering
\includegraphics[width=0.85\columnwidth]{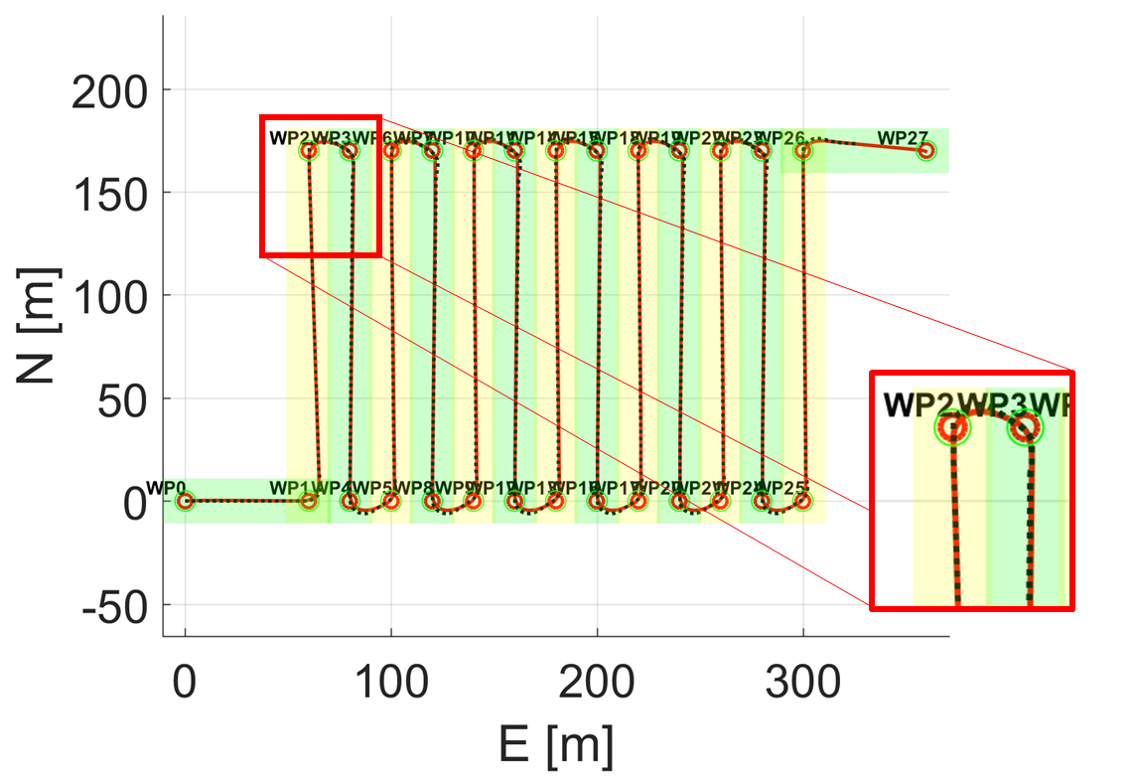}
\caption{Comparison among (red line) and improved HIL (dotted black line) trajectories.}
\label{fig:f103}
\end{figure}

In order to improve the effectiveness of the TRMPC approach in path tracking for HIL simulations, the state weight matrix value $Q_{lat}(4,4)$ related to roll angle $\phi$, has been slightly modified. Changing this value of the state weight matrix, we can notice how the SIL and improved HIL trajectories are almost perfectly overlapped (Figure \ref{fig:f103}). 


\section{Conclusions}
\label{conc}
In this paper the application of Unmanned Aerial Vehicles (UAVs) for crop monitoring is described, to improve the farm efficiency and productivity, reducing the cost and increasing the mapping flexibility. In this research, a guidance algorithm and a robust control system are combined to guarantee the robustness of the system to additive noise (i.e. wind disturbance) and uncertainties (i.e. model parameter variations). A waypoint-grid on a paddy field is verified by hardware-in-the loop tests, to prove the real-time feasibility of the proposed approach and to maximize the selected sensor performance. Good results are obtained, increasing the sensor coverage and guaranteeing good stability performance of the platform. This research proves: (i) the advantages of using UAVs in precision farming, and (ii) the repeatability of the monitoring. Moreover, thanks to the controller robustness, a re-tuning of the software is not required, thus this system is applicable as it is to a fleet of UAVs.
\section*{Acknowledgements}
The authors would like to thank Dr. Gianluca Ristorto, Senior Engineer at MAVTech srl, for the insightful discussion on precision farming scenario and on the hardware in the loop tests.
\nocite{*}
\bibliographystyle{IEEEtran}
\bibliography{BIBLIO}

\begin{thebibliography}{10}
\providecommand{\url}[1]{#1}
\csname url@samestyle\endcsname
\providecommand{\newblock}{\relax}
\providecommand{\bibinfo}[2]{#2}
\providecommand{\BIBentrySTDinterwordspacing}{\spaceskip=0pt\relax}
\providecommand{\BIBentryALTinterwordstretchfactor}{4}
\providecommand{\BIBentryALTinterwordspacing}{\spaceskip=\fontdimen2\font plus
\BIBentryALTinterwordstretchfactor\fontdimen3\font minus
  \fontdimen4\font\relax}
\providecommand{\BIBforeignlanguage}[2]{{%
\expandafter\ifx\csname l@#1\endcsname\relax
\typeout{** WARNING: IEEEtran.bst: No hyphenation pattern has been}%
\typeout{** loaded for the language `#1'. Using the pattern for}%
\typeout{** the default language instead.}%
\else
\language=\csname l@#1\endcsname
\fi
#2}}
\providecommand{\BIBdecl}{\relax}
\BIBdecl

\bibitem{pederi}
Y.~A. Pederi and H.~S. Cheporniuk, ``Unmanned aerial vehicles and new
  technological methods of monitoring and crop protection in precision
  agriculture,'' in \emph{2015 IEEE International Conference Actual Problems of
  Unmanned Aerial Vehicles Developments (APUAVD)}, Oct 2015, pp. 298--301.

\bibitem{ni}
J.~Ni, L.~Yao, J.~Zhang, W.~Cao, Y.~Zhu, and X.~Tai, ``Development of an
  unmanned aerial vehicle-borne crop-growth monitoring system,''
  \emph{Sensors}, vol.~17, no.~3, p. 502, 2017.

\bibitem{ghazal}
\BIBentryALTinterwordspacing
M.~Ghazal, Y.~A. Khalil, and H.~Hajjdiab, ``Uav-based remote sensing for
  vegetation cover estimation using ndvi imagery and level sets method,'' in
  \emph{2015 IEEE International Symposium on Signal Processing and Information
  Technology (ISSPIT)}, vol.~00, Dec. 2015, pp. 332--337. [Online]. Available:
  \url{doi.ieeecomputersociety.org/10.1109/ISSPIT.2015.7394354}
\BIBentrySTDinterwordspacing

\bibitem{voug}
S.~Vougioukas, K.~Arvanitis, and N.~Sigrimis, ``A non linear model predictive
  tracking controller for agricultural vehicles,'' in \emph{2007 European
  Control Conference (ECC)}, July 2007, pp. 4937--4943.

\bibitem{cannon}
B.~Kouvaritakis and M.~Cannon, \emph{Model Predictive Control: Classical,
  Robust and Stochastic}.\hskip 1em plus 0.5em minus 0.4em\relax Advanced
  Textbooks in Control and Signal Processing, Springer, 2015.

\bibitem{raw}
D.~Q. Mayne and J.~B. Rawlings, \emph{Model Predictive Control: Theory and
  Design}.\hskip 1em plus 0.5em minus 0.4em\relax Nob Hill Publishing, 2009.

\bibitem{barros}
S.~R.~B. dos Santos, C.~L.~N. J{\'u}nior, S.~N.~G. Junior, A.~Bittar, and
  N.~M.~F. de~Oliveira, ``Experimental framework for evaluation of guidance and
  control algorithms for uavs,'' in \emph{Proceedings of the 21st Brazilian
  Congress of Mechanical Engineering}, 2011.

\bibitem{smile}
{MAVTech srl}, ``{Smile Project},''
  \url{//www.mavtech.eu/site/assets/files/1423/progetto_smile_eng.pdf}, 2014.

\bibitem{cap2012}
\BIBentryALTinterwordspacing
E.~Capello, G.~Guglieri, P.~Marguerettaz, and F.~Quagliotti, ``Preliminary
  assessment of flying and handling qualities for mini-uavs,'' \emph{Journal of
  Intelligent {\&} Robotic Systems}, vol.~65, no.~1, pp. 43--61, Jan 2012.
  [Online]. Available: \url{https://doi.org/10.1007/s10846-011-9565-5}
\BIBentrySTDinterwordspacing

\bibitem{cap13}
\BIBentryALTinterwordspacing
E.~Capello, G.~Guglieri, F.~Quagliotti, and D.~Sartori, ``Design and validation
  of an l1 adaptive controller for mini-uav autopilot,'' \emph{Journal of
  Intelligent {\&} Robotic Systems}, vol.~69, no.~1, pp. 109--118, Jan 2013.
  [Online]. Available: \url{https://doi.org/10.1007/s10846-012-9717-2}
\BIBentrySTDinterwordspacing

\bibitem{cap2008}
E.~Capello, G.~Guglieri, and F.~B. Quagliotti, ``A software tool for mission
  design and autopilot integration: an application to micro aerial vehicles,''
  in \emph{Proceedings of the 2008 Summer Computer Simulation
  Conference}.\hskip 1em plus 0.5em minus 0.4em\relax Society for Modeling \&
  Simulation International, 2008, p.~9.

\bibitem{stevens}
B.~Stevens and F.~Lewis, \emph{Aircraft Control and Simulation}.\hskip 1em plus
  0.5em minus 0.4em\relax New York: John Wiley and Sons, 2003.

\bibitem{etk}
B.~Etkin and L.~Reid, \emph{Dynamics of Flight: Stability and Control}.\hskip
  1em plus 0.5em minus 0.4em\relax New York: John Wiley and Sons, 1996.

\bibitem{martins}
\BIBentryALTinterwordspacing
G.~Martins, A.~Moses, M.~J. Rutherford, and K.~P. Valavanis, ``Enabling
  intelligent unmanned vehicles through xmos technology,'' \emph{The Journal of
  Defense Modeling and Simulation}, vol.~9, no.~1, pp. 71--82, 2012. [Online].
  Available: \url{https://doi.org/10.1177/1548512910388197}
\BIBentrySTDinterwordspacing

\bibitem{cap2014}
\BIBentryALTinterwordspacing
E.~Capello, G.~Guglieri, and G.~Ristorto, ``Guidance and control algorithms for
  mini uav autopilots,'' \emph{Aircraft Engineering and Aerospace Technology},
  vol.~89, no.~1, pp. 133--144, 2017. [Online]. Available:
  \url{https://doi.org/10.1108/AEAT-10-2014-0161}
\BIBentrySTDinterwordspacing

\bibitem{mascarello2017feasibility}
L.~N. Mascarello, F.~Quagliotti, and G.~Ristorto, ``A feasibility study of an
  harmless tiltrotor for smart farming applications,'' in \emph{Unmanned
  Aircraft Systems (ICUAS), 2017 International Conference on}.\hskip 1em plus
  0.5em minus 0.4em\relax IEEE, 2017, pp. 1631--1639.

\end{thebibliography}

\end{document}